\documentclass{aastex631}

\usepackage{amsmath}
\usepackage{xcolor}

\shorttitle{Exotic compact objects with two dark matter fluids}
\shortauthors{Cassing, Brisebois, Azeem, Schaffner-Bielich}

\graphicspath{{./}{figures/}}

\begin{document}

\title{Exotic Compact Objects with Two Dark Matter Fluids}

\author[0000-0002-0244-2983]{Marie Cassing}
\affiliation{Institut f{\"u}r Theoretische Physik, 
Max-von-Laue-Straße 1, 60438 Frankfurt am Main, Germany}

\author[0000-0002-2909-0877]{Alexander Brisebois}
\altaffiliation{Present Affiliation: Department of Physics and Astronomy, 
University of Waterloo, Waterloo, Ontario N2L 3G1, Canada }
\affiliation{Division of Engineering Science,
University of Toronto, Toronto, Ontario M5S 2E4, Canada} 

\author[0000-0002-2796-9474]{Muhammad Azeem}
\affiliation{Department of Physics and Astronomy, 
York University, Toronto, Ontario M3J 1P3, Canada}

\author[0000-0002-0079-6841]{J{\"u}rgen Schaffner-Bielich}
\affiliation{Institut f{\"u}r Theoretische Physik, 
Max-von-Laue-Straße 1, 60438 Frankfurt am Main, Germany}

\begin{abstract}
The generic properties of compact objects made of two different fluids of dark matter are studied 
in a scale invariant approach. 
We investigate compact objects with a core-shell structure, where the two fluids are separated, 
and with mixed dark matter components, where both dark matter fluids are immersed within each other.
The constellations considered are combinations of incompressible fluids, free and interacting Fermi gases, 
and equations of state with a vacuum term, i.e. self-bound dark matter. We find novel features in the 
mass-radius relations for combined dark matter compact objects which distinguishes them from compact objects with a single
dark matter fluid and compact stars made of ordinary baryonic matter, as white dwarfs, neutron stars and quark stars. 
The maximum compactness of certain combined dark matter stars can reach values up to
the causality limit for compact stars but not beyond that limit if causality of the dark matter fluids is
ensured.
\end{abstract}

%%%%%%%%%%%%%%%%%%%%%%%%%%%%%%%%%%%%%%%%%%%%%%%%%%%%%%%%%%%%%%%%%%%%%%%%%%%%%%

\section{Introduction} \label{sec:intro}

The phenomena of dark matter are so far unexplained features from cosmological observations
and astrophysical data from the small and large scale structure of the universe. 
Possible candidates of dark matter are weakly interacting massive particles WIMPs, which are fermions,
or axions and axion-like particles.
However, the standard cold dark matter (CDM) model of collisionless free particles is in tension with the
small scale structure observations of galaxies as e.g.\ the core-cusp problem and the satellite problem.
Selfinteracting dark matter can provide a possible explanation of these observations, see
\citep{Tulin:2017ara}.

If dark matter consists of selfinteracting particles they can form spheres of a fluid which are bound by gravity, 
i.e.\ compact stars of dark matter. Boson stars made of bosonic dark matter particles
have been discussed for a long time, first by a noninteracting scalar field \citep{Wheeler:1955zz,Kaup:1968zz,Ruffini:1969qy} and then by introducing selfinteractions \citep{Colpi:1986ye,Eby:2015hsq}
or a solitonic potential \citep{Lee:1986tr,Lee:1986ts,Friedberg:1986tp,Friedberg:1986tq}.
The case for fermions, the properties of compact stars made of selfinteracting fermionic dark matter compact stars, have been investigated in addition, see e.g.\ \citep{Narain:2006kx,Kouvaris:2015rea,Gresham:2018rqo}.
Stable compact stars which are bound without gravity can be generated by a specific form of the equation of state where the pressure vanishes at a nonvanishing energy density. These are so called self-bound stars which have been discussed for absolutely stable strange quark matter \citep{Witten:1984rs,Alcock:1986hz,Haensel:1986qb}, for a review see \citep{Weber:2004kj}. As shown in \citep{Chang:2018bgx} compact stars made of interacting dark matter particles can be produced by cosmological structure formation on various different scales depending on the interaction strength and masses of the dark matter particles.

Dark matter can change the properties of neutron stars and white dwarfs. Thereby, neutron stars and white dwarfs can serve as
probes for dark matter and have the potential to constrain the properties of dark matter
\citep{Bertone:2007ae,Kouvaris:2007ay,deLavallaz:2010wp,McCullough:2010ai,Kouvaris:2010vv,Kouvaris:2010jy,Perez-Garcia:2011tqq,McDermott:2011jp,Bell:2013xk,Garani:2018kkd}. The properties of hypothetical quark stars will be also altered when 
effects from a dark matter fluid are considered \citep{Mukhopadhyay:2015xhs}.
The mass-radius relation of neutron stars can significantly change if there is dark matter present.
The modification of the global structure of neutron stars has been studied in detail for bosonic dark matter and
fermionic dark matter with and without selfinteractions, see
\citep{Sandin:2008db,Ciarcelluti:2010ji,Leung:2011zz,Li:2012ii,Xiang:2013xwa,Mukhopadhyay:2016dsg,Ellis:2018bkr,McKeen:2018xwc,Baym:2018ljz,Motta:2018rxp,Ivanytskyi:2019wxd,Husain:2022bxl} and can even lead to exotic compact objects as dark compact planets \citep{Tolos:2015qra}.
Present and future gravitational wave detectors have the potential to detect the possible presence of dark matter in
merging neutron stars and to constrain the properties of dark matter, as its mass and its selfinteraction strength
\citep{Ellis:2017jgp,Nelson:2018xtr,Horowitz:2019aim,Bauswein:2020kor,Dengler:2021qcq,Karkevandi:2021ygv}

The measurements of gravitational waves from binary mergers of compact objects 
has the potential to observe and study exotic compact objects made of dark matter
\citep{Cardoso:2016oxy,Maselli:2017vfi,Maselli:2017vfi,Maselli:2017tfq,Mark:2017dnq,Gresham:2018rqo,Toubiana:2020lzd,Wystub:2021qrn}
and distinguish them from neutron stars and black holes \citep{Sennett:2017etc}. 
Numerical simulations of the gravitational wave spectrum for neutron stars with dark matter have been put forward just recently by \citep{Emma:2022xjs}.
The LIGO-Virgo collaboration has published a catalog of gravitational wave sources from binary mergers
from their third run, GWTC-3 \citep{LIGOScientific:2021djp}.
The catalog exhibits several masses of compact objects which are lying in the mass gap between the mass of the lightest black hole, set to $5M_\odot$, and the most massive neutron stars with a maximum mass of about $2M_\odot$ from pulsar data \citep{NANOGrav:2019jur,Fonseca:2021wxt}.
A unique identification of a neutron star merger has been so far only be possible for 
GW170817 by the observation of a jet and the optical afterglow in the form of a kilonova 
\citep{LIGOScientific:2020zkf}. Other neutron star-neutron star or neutron star-black hole merger candidates are classified by their masses. It is not clear if these gravitational wave events emerge from compact objects involving ordinary neutron stars. The gravitational wave event GW190814, for example, involves a compact object with a mass of $2.6M_\odot$ \citep{LIGOScientific:2020zkf}, well above the most massive neutron star known at present.

Dark matter can also be present in more sophisticated forms, as in the form of multi-component dark matter which arises naturally in models beyond the standard model \citep{Zurek:2008qg}. There could be different kinds of particles, 
for example in the form of composite dark matter \citep{Khlopov:2007ic,Khlopov:2008ty}.
Recently, the exploration of the properties of compact stars made of two different fluids of dark matter has been put forward by studying dark white dwarfs \citep{Ryan:2022hku} and exotic cores \citep{Zollner:2022dst}. 
In this work we extend those studies of the properties of spheres of dark matter for several cases of
two different kinds of fluids of dark matter. To keep our discussion more general, we use scaling
relations so that our results can be adopted for arbitrary scales, be it masses, vacuum energies or 
interaction strengths.
In Section \ref{sec:theo} the Tolman-Oppenheimer-Volkoff equation for gravitating spheres of fluids is introduced in scale invariant form and its scaling relations are discussed. The generic equations of state for an interacting gas of fermions, self-bound matter and the analytic solution for an incompressible fluid are summarized then.
In Section \ref{sec:results} the results of our investigations are presented for the following specific cases which can be seen in table \ref{Cases}:
i) Two incompressible fluids,
ii) incompressible fluid core and an interacting Fermi gas shell,
iii) interacting Fermi gas core and a free Fermi gas shell,
iv) self-bound matter core and an interacting Fermi gas shell,
and v) two homogeneously mixed interacting Fermi fluids. 
The results of our study are summarized in Section~\ref{sec:conclusions}. 

\begin{table}
\begin{center} \label{Cases}
   \begin{tabular}{ | c | c | l | p{5cm} }
     \hline
     Model & Section & Description \\ \hline
     1 & 3.1 & Two incompressible fluids \\ \hline
     2 & 3.2 & Incompressible fluid core and an interacting Fermi gas shell \\ \hline
     3 & 3.3 & Interacting Fermi gas core and a free Fermi gas shell \\ \hline
     4 & 3.4 & Self-bound matter core and an interacting Fermi gas shell \\ \hline
     5 & 3.5 & Star of two homogeneously mixed interacting Fermi fluids \\ 
     \hline 
   \end{tabular} 
\end{center}
\caption{The different models of exotic compact objects consisting of two fluids of dark matter studied in this work}
\end{table}

%%%%%%%%%%%%%%%%%%%%%%%%%%%%%%%%%%%%%%%%%%%%%%%%%%%%%%%%%%%%%%%%%%%%%%%%%%%%%%

\section{Theoretical background} \label{sec:theo}

For a perfect fluid, the equations for hydrostatic equilibrium in general relativity are denoted as the Tolman–Oppenheimer–Volkoff (TOV) equations and can be derived by solving the Einstein-equations for a spherically symmetric and static  metric. The latter are given by the following differential equations:
\begin{align}
\frac{dP}{dr} &= \frac{-Gm}{r^2} \rho \left( 1 + \frac{P}{\rho} \right) \left(1+\frac{4 \pi r^3 P}{m} \right) \left(1-\frac{2Gm}{r} \right)^{-1} \label{tov_p} \\
\frac{dm}{dr} &= 4 \pi r^2 \rho , \label{tov_m}
\end{align}
where $P$ and $\rho$ denote the pressure and energy density, $m$ and $r$ the mass and radius, respectively, and $G$ is the gravitational constants. Note that throughout the paper we are using natural units, i.e. $\hbar = c = 1$.
For a given equation of state, which provides a relation between $P$ and
$\rho$,  we can solve the above set of equations employing appropriate initial and boundary conditions. We denote the total radius of a star by $R$,  which is found using the condition that the pressure vanishes at the surface of the star ($P(R)=0$). The mass $M(r=0)$ must be zero at $r = 0$ and $M(R)$ gives the total mass of the star at $r = R$. In the model cases investigated here these quantities are found numerically by solving the differential equations (\ref{tov_p}) and (\ref{tov_m}) for a star containing different types of matter and therefore different equations of state.

\subsection{ Scaling the TOV equation }

The TOV equations contain a dimensional quantity $G$, which can expressed in terms of the Planck mass ($G = M_p^{-2}) $. To transform the TOV equation to scale independent variables we use the rescaling as introduced in \citep{Narain:2006kx}, see also \citep{Schaffner-Bielich:2020psc}
(throughout the paper we will denote dimensional quantities with a prime).
The rescaling factor $\epsilon_0$ can be calculated for a given equation of state. The dimensionless form of the equations facilitates the computational solutions since one does not have to care about units,
\begin{equation}
P = \epsilon_0 P' , \qquad \epsilon = \epsilon_0 \epsilon'  \, .
\end{equation}
Pressure and energy-density have the same units and thus can be rescaled by the constant $\epsilon_0$ (for a Fermi gas of mass $m_f$ we use $\epsilon_0 = m_f^4 $). Then the radius and the mass also have to be rescaled:
\begin{equation}
r=a r' , \qquad   M=b M'
\label{dimless1}
\end{equation}
Plugging the new quantities into the TOV-equation we obtain a dimensionless TOV-equation and relations for the rescaling factors $a$ and $b$ which we can use below to obtain these quantities in physical units,
\begin{equation}
a= (G \epsilon_0)^{-1/2}, \qquad  b= (G^3 \epsilon_0)^{-1/2}  \, .
\label{dimless2}
\end{equation}
We will now discuss the different types of equations of state (EOS) used as input for the TOV-equation.

\subsection{Constant Density Solution}

An analytical solution for the above set of TOV-equations for the interior of a star $r\leq R$ can be obtained for  an incompressible fluid, i.e. for constant density $\rho(r) = \rho_0$. In this case the solution for Eq.~(\ref{tov_m}), for $r \leq R $, is:
\begin{equation}
M(r)=\frac{4\pi}{3} r^3 \rho_0  . \label{mass_incomp_fluid}
\end{equation}
With (\ref{mass_incomp_fluid}) and the boundary condition $P(r=R) = 0$  the solution of the TOV-equation is given by  the Schwarzschild interior solution \citep{Schwarzschild:1916b} (see also \citep{Misner:1974qy}) :
\begin{equation}
P(r)= \rho_0 \left[ \frac{R\sqrt{R-2GM}-\sqrt{R^3-2GMr^2}}{\sqrt{R^3-2GMr^2}-3R\sqrt{R-2GM}}\right] .
\end{equation}
The  central pressure (at $r=0$) is related to the mass $M$ by
\begin{equation}
P(r=0)= \rho_0 \left[ \frac{\sqrt{1-\frac{2GM}{R}}-1}{1-3\sqrt{1-\frac{2GM}{R}}}\right] . \label{central_p}
\end{equation}
We define the compactness of a star by the mass to radius ratio,
\begin{align}
C = \frac{GM}{R}.
\end{align}
Note that the central pressure, eq.~(\ref{central_p}), diverges for $R \rightarrow \frac{9GM}{4}$,  which is known as the Buchdahl-limit \citep{Buchdahl:1959zz} giving the maximum compactness $C_{max}=\frac{4}{9}$.
Objects with a compactness larger than the Buchdahl limit will be black holes. Another constraint on the maximal mass and compactness is the causality limit $C_s \approx 3/8$ \citep{Schaffner-Bielich:2020psc}.
It arises from the fact that the speed of sound in the star has to be smaller then the speed of light $c$ to fulfill causality and thus $c_s^2 \leq 1$, i.e.\ by setting $P(r=0)=\rho$. The speed of sound squared $c^2_s$ in a medium  is defined by the derivative of the pressure with respect to the energy density $\rho$ at constant entropy $S$,
\begin{equation}
c_s^2 = \left( \frac{\partial P}{\partial \rho} \right)_S .
\end{equation}
Considering a polytropic equation of state the polytropic index $\Gamma=2$ will lead to the stiffest possible EoS and the limit $P=\rho $ with $c_s^2=1$.
The speed of sound for a fermionic type of matter -- which is used in four of our models -- is $c_s = 1/ \sqrt{3}$  at high densities, thus obeying causality.

In most cases studied in this paper, we consider a
core-shell-structure of the star and a fluid density profile for a star containing two different fluids in the core and the shell. 
We investigate stars with a core of fluid density $\rho_0$ and dimensionless core radius $r_c$ and a shell of fluid density $\rho_1$.  
Then the density profile reads as follows:
\begin{align}
\rho(r)=
        \begin{cases}
            \rho_0, & 0 \leq r \leq r_c \\
            \rho_1, & r_c \leq r \leq R, \\
            0 & \text{else}
        \end{cases} .
\end{align}
Here the density in the core $ \rho_0 $ and the density in the shell $\rho_1$ can be a function of $P$ depending on the equation of state for each fluid. For such a profile there is no longer an analytical solution for the TOV equations (\ref{tov_p}) and (\ref{tov_m}) and the latter have to be solved  numerically (e.g.\ by the Runge-Kutta method). 
The construction of these stars is done such that the pressure is continuous and smooth at the transition radius $r_c$. Due to the different equations of state in the core and the shell, the energy density of the second matter type can be lower for the same pressure. 
This results in a first order phase transition and a jump in the density at the core radius $r_c$. Such a behaviour is present in model 1 when transitioning from a higher constant density $\rho_0$ of the incompressible fluid to a lower constant density of the incompressible fluid $\rho_1$ in the shell. 
This density profile has a step shape. A first order phase transition can also occur in model 2 transitioning from the incompressible fluid as well as in model 4 transitioning from the self-bound matter to lower density of the interacting fermi gas  . 

\subsection{Fermi gas with interactions}

\noindent We use the same equation of state as in Ref. \citep{Narain:2006kx} for a  Fermi gas of mass $m_f$ with interactions at zero temperature. The EoS for an interacting Fermi gas is given as an explicit expression of $P$ and $\rho$:
\begin{eqnarray}
\rho &=& \frac{g}{2\pi^2}\int_0^{p_F} dp E p^2 + \left(\frac{1}{3}\pi^2 \right)^2 y^2 p_F^6 = m_f^4 \rho^\prime
\label{fermi_int_energy_dens_eq}\\
P &=& \frac{g}{2\pi^2}\int_0^{p_F} dp \frac{p^4}{3 E} + \left(\frac{1}{3}\pi^2 \right)^2 y^2 p_F^6  = m_f^4 P^\prime \label{fermi_int_press_eq}
\end{eqnarray}
with $y = m_f/m_I$ being the interaction strength, where $m_I$ is the scale of the interaction, and $p_F$ the Fermi momentum of the particles. The spin-factor for $s= 1/2$ fermions, $g= 2$. Note that when $y=0 $ we get the non-interacting Fermi gas i.e free Fermi gas.

\noindent For comparison, the mass of a neutron is $m_n = 0.938 \,  GeV/c^2$ and we use the neutron mass or fractions of it to describe the different types of matter. 
Using $m_1=m_n$ and some type of dark matter with $m_2 = 0.5 m_1$ would lead to a configuration similar to a white dwarf.
For the rescaling of the interacting Fermi gas we calculate the rescaling factor using equations (\ref{dimless1}) and (\ref{dimless2}) to $\epsilon_0 = m_f^4$ for dimensionless units.

\subsection{Conformal equation of state with vacuum energy}

Models with an non-vanishing vacuum energy are used to describe self-bound stars. 
A well known example of such a vacuum model is the MIT-bag model which is often used for the description of cold  quark stars, see \citep{Weber:2004kj} for a review.
The MIT-bag model is a phenomenological model for mimicking quark confinement and was introduced to describe hadron properties in terms of a bag of quarks with a bag constant $B$. 
The vacuum energy density in this model equals to 4 times the bag constant.
\begin{equation}
\rho_{vac}=4B , \quad \text{ with } B^{1/4} \approx 145 \,  \text{MeV} .
\end{equation}
It allows self-bound stars to be stabilized by itself as without gravity.
The equation of state for the MIT bag model is
\begin{equation}
P = \frac{1}{3} (\rho - \rho_{vac}) \, .
\label{eq:vacuumEOS}
\end{equation}
Such an equation of state is used in the fourth model case of our studies for the description of a core of self-bound matter.
It is related to the constant speed of sound parameterization used for describing quark matter in compact stars
\citep{Alford:2013aca} with a speed of sound of $c_s^2=1/3$ and can be motivated from perturbative QCD calculations
for quark matter at high density \citep{Fraga:2001id}. We will use eq.~(\ref{eq:vacuumEOS}) generically as an EOS for self-bound dark matter below. We will set $\rho_{vac}$ as the scaling variable $\epsilon_0$, so that the results can be rescaled accordingly without the need to specify the vacuum energy scale explicitly. The form of the EOS given in eq.~(\ref{eq:vacuumEOS}) emerges also from conformal field theories and is also known as the conformal EOS.

%%%%%%%%%%%%%%%%%%%%%%%%%%%%%%%%%%%%%%%%%%%%%%%%%%%%%%%%%%%%%%%%%%%%%%%%%%%%%%

\section{Combined dark matter compact objects} \label{sec:results}

%##########################################################################
%++++++++++++++++++++++++++++++ incompressible fluids++++++++++++++++++++++

\subsection{Incompressible fluid stars} \label{sec:incompr}

\begin{figure}
	\centering
		\includegraphics[width=0.45\textwidth]{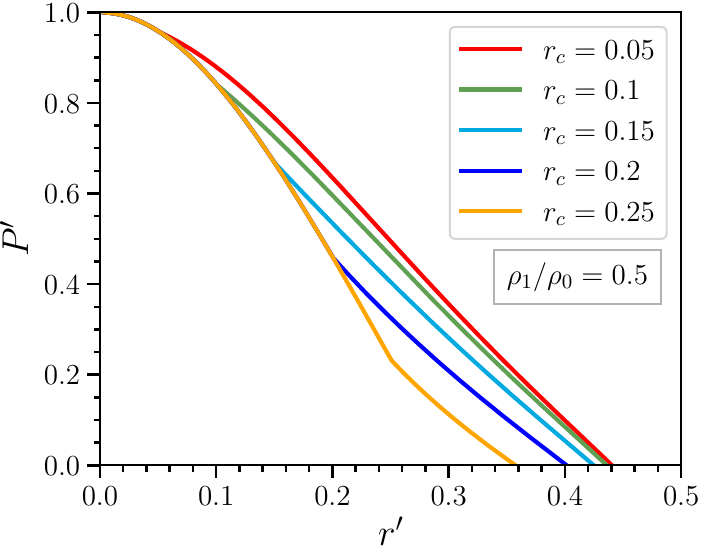}
		\includegraphics[width=0.45\textwidth]{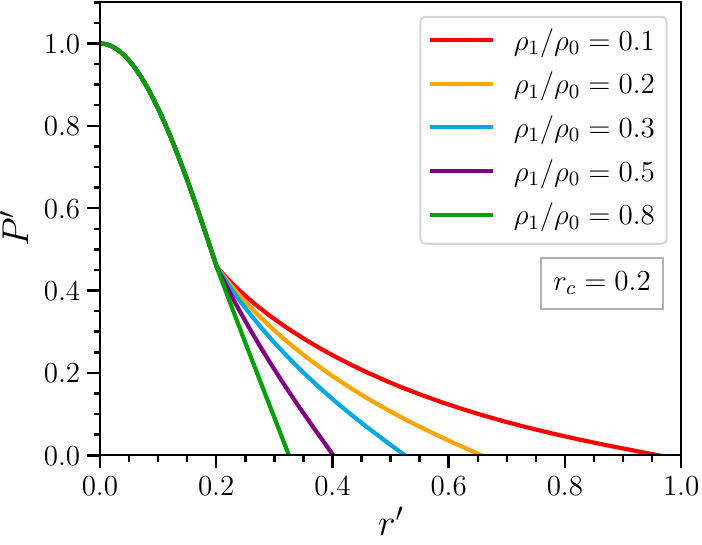}
	\caption{The dimensionless pressure $P'$ of a star with a 2-step density profile and a fixed central pressure $P'_c=1$ displayed as a function of the radial coordinate $r'$ for two cases: i) for varying core radius $r_c$ with fixed ratio $\rho_1/\rho_0$ (left)\, and ii) for varying density ratios $\rho_1/\rho_0$ with fixed $r_c$ (right).}
	\label{star_incomp_P0_fixed}
\end{figure}

\begin{figure}
\centering
\includegraphics[width=0.45\textwidth]{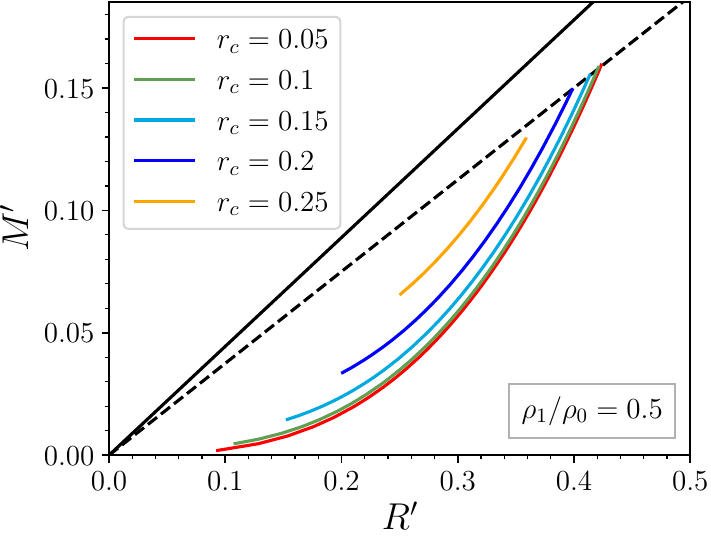}
\includegraphics[width=0.45\textwidth]{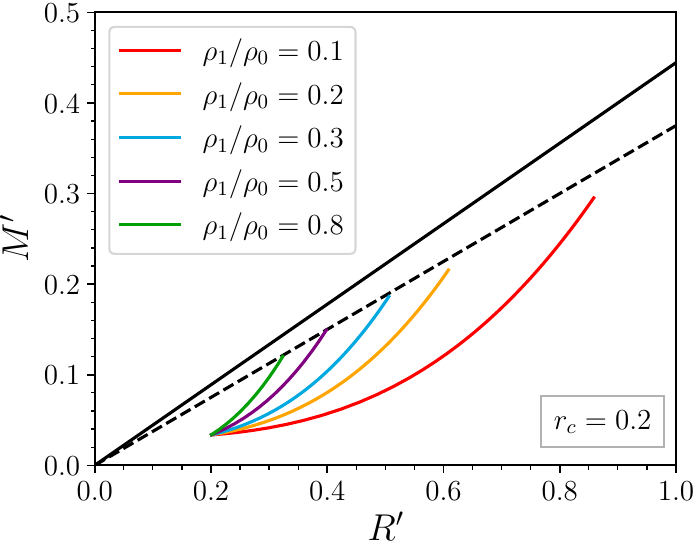}
\caption{The mass-radius-relation for different central pressures $0 \leq P_c \leq 1$ for a star with a 2-step density profile containing two incompressible fluids with a density-ratio of $\rho_1 / \rho_0 = 0.5 $ and varying core-radii (left), and with a core-radius of $r_c = 0.2$ and varying density-ratios $\rho_1 / \rho_0$ (right).  The dashed and solid black lines represent the causality and Buchdahl limits, respectively.}
\label{star_incomp_P0_var}
\end{figure}

In this case we examine a star consisting of two layers of incompressible fluids with densities $\rho_0$ and $\rho_1$ for the core and shell, respectively.  The density changes at a chosen critical radius denoted by $r_c$.  Two separate cases are explored in this section (Fig. \ref{star_incomp_P0_fixed}): i) the ratio between the constant densities is fixed to $\rho_1/\rho_0 = 0.5$ while the core-radius $r_c$ is varied; ii) the critical radius is fixed to $r_c = 0.2$ while the ratio of the densities $\rho_1/\rho_0$ is varied. To demonstrate the shape of a single pressure profile throughout the star, the central pressure is fixed at $P_c = 1$ and the profiles for the two different cases are presented in Fig. \ref{star_incomp_P0_fixed}.

The pressure profiles for case i)  with a fixed density-ratio and varying $r_c$ is shown on the left of Fig. \ref{star_incomp_P0_fixed} and are continuous throughout the star. The slope of the pressure deviates from that of a single fluid star at the chosen values of $r_c$ forming a kink.
In case ii) with fixed $r_c$ and varying density-ratio $\rho_1/\rho_0$ (r.h.s. of the figure) the slope of the pressure similarly deviates at $r = r_c$, while the magnitude of the slope decreases with dropping ratios $\rho_1/\rho_0$.

In Fig. \ref{star_incomp_P0_var} we display the relation between the total mass $M'$ and the total radius $R'$ of the star for the two cases i) and ii), but for different values of the central pressure $P'_c$.
For some combinations of $r_c$ and $\rho_1/\rho_0$ we reach the causality limit, which results in very compact configurations.

From the mass-radius relation of the $r_c$-varying case i)  (left of Fig. \ref{star_incomp_P0_var}),  lower values of $r_c$ result in a higher compactness for configurations with high central pressures $P_c$ (some reaching the causality limit). In case ii) (right of Fig. \ref{star_incomp_P0_var}) with varying ratio $\rho_1/\rho_0$, a higher $\rho_1/\rho_0$ will result in a star with a higher compactness, which implies that configurations with larger shell density are more likely to reach the causality limit. One also notices  that configurations with a small density ratio reach higher maximal masses and larger radii for high central pressure.

%##########################################################################
%+++++++++++++++++++++++++++++ Incompr + fermi+++++++++++++++++++++++++++++

\subsection{Incompressible fluid and Fermi gas star} \label{sec:incompr_fermi}

\begin{figure}
\centering
\includegraphics[width=0.45\textwidth]{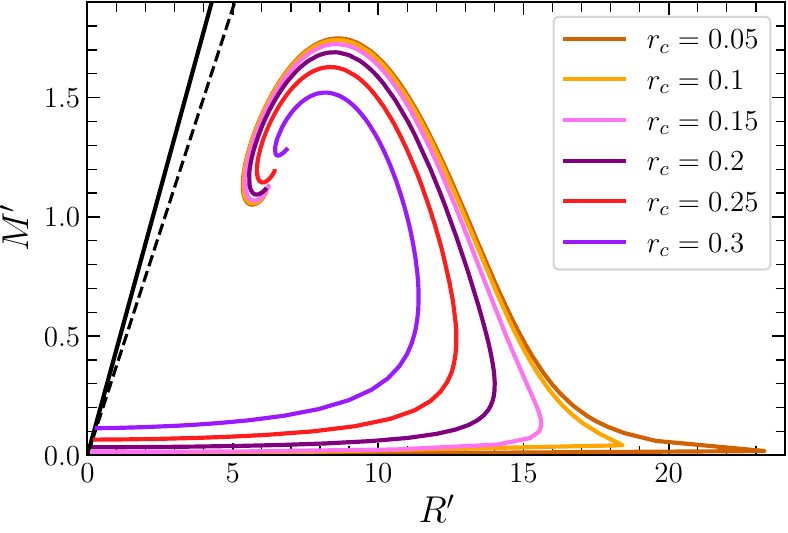}
\includegraphics[width=0.45\textwidth]{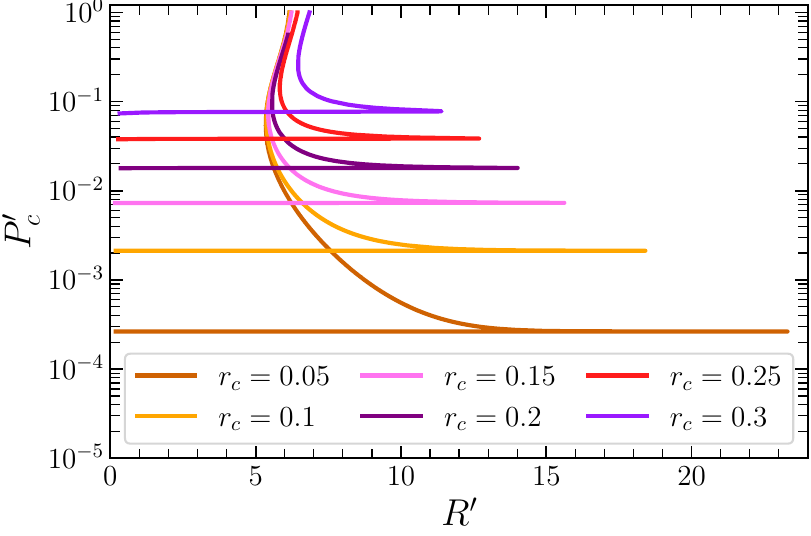}
\caption{The dimensionless maximal mass $M'$ as a function of the radius $R'$ for different initial values of the central pressure $P'_c$ (left) and the dimensionless central pressure $P'_c$ as a function of the radius of the star $R'$ (right) for a star combined out of an incompressible fluid up to different core radii $0.1 \leq r_c \leq 0.3$ and a Fermi gas with interactions ($y=10$) as a shell. Again the dashed and solid lines represent the causality and Buchdahl limit.} 
\label{incomp_fermi_plot}
\end{figure}

In this model we consider an incompressible fluid core and a shell with an interacting Fermi gas. In our example the mass of the fermion particles is $m_f$, which might be some kind of dark matter with interaction strength $y=10$. 
In the core the energy density of the incompressible fluid is set to be equal to the maximal possible value of the central pressure $P_{c,max}=\epsilon_0=1$. The TOV equations then are solved for different values of the central pressure, and for small $P_c$ we would see a first order phase transition in the energy density. An analytical solution of the TOV equations can be found for the core of the star. The mass as a function of the radius then can be calculated as follows.
\begin{equation}
M=\frac{4\pi}{3}\epsilon_0 r^3 \, . \label{M_incomp_analyt}
\end{equation}
The compactness at the core should not be larger than the Buchdahl limit $M_c/r_c \leq 4/9$ and inserting eq.~(\ref{M_incomp_analyt}) we get a limit on the possible core radii in dimensionless units $r'_c \leq \sqrt{1/(3\pi)} \approx 0.33$. Another constraint on the core radius comes from the causality limit $M_c/r_c \leq 3/8$. In dimensionless units the core radius should be smaller than $r'_c \leq \sqrt{9/(32\pi)}\approx 0.3$.
An analytical solution for the pressure in the core of the star can also be obtained in the following form:
\begin{equation}
 P_1(r)=\epsilon \left[ \frac{\epsilon_0 + P_c -(3P_c + \epsilon_0)\sqrt{1-\frac{8\pi}{3}\epsilon_0 r^2}}{(\epsilon_0 + 3 P_c)\sqrt{1-\frac{8\pi}{3}\epsilon_0 r^2} - 3(\epsilon_0 + P_c)} \right]  \, .
\end{equation}
Now for a specific value of the central pressure $P_c$ and a selected core radius $r_c$ the pressure at the core radius should be continuous and we would start to solve for the interacting Fermi gas starting from the pressure $P_1(r_c)$.
In Fig. \ref{incomp_fermi_plot} we see the typical shape of the mass-radius-curves. We obtain a swirl and an unstable branch for large central pressures due to the behaviour of the interacting Fermi gas. To the right of the maximal mass $M_{TOV}$ we drop to the stable branch with smaller central pressures. It can be seen that a larger core radius and therefore a larger amount of the incompressible fluid leads to a smaller maximal mass $M_{TOV}$. It can be explained such that interactions in the Fermi gas allow for more compact configurations and increasing the core radius decreases the amount of the surrounding Fermi gas. For low pressures the mass has a turning point and curves to the left in the direction of small radii $R$. This branch appears due to the presence of the incompressible fluid core which results in the Fermi solution for larger central pressures. For larger core radii this transition happens smoothly and for smaller core-radii we find a sharper turning point.
On the right hand side of Fig.~\ref{incomp_fermi_plot} we see that for some small values of the central pressure $P_c$ the total radius $R$ increases very fast, such that we obtain an almost horizontal line for the central pressure.
Therefore for small central pressures we first obtain small total radii and then large total radii of the star. At the point where the Fermi gas dominates over the incompressible fluid and the mass-curve has a turning point the total radius of the star $R$ decreases for increasing $P_c$. It then turns up and  back to a bit larger value of $R$ staying almost constant afterwards for increasing $P_c$.

%##########################################################################
%++++++++++++++++++++++++++++++ 2 Fermi gases+++++++++++++++++++++++++++++

\subsection{Two Fermi gases star} \label{sec:fermi2}

\begin{figure}
	\centering
		\includegraphics[width=0.45\textwidth]{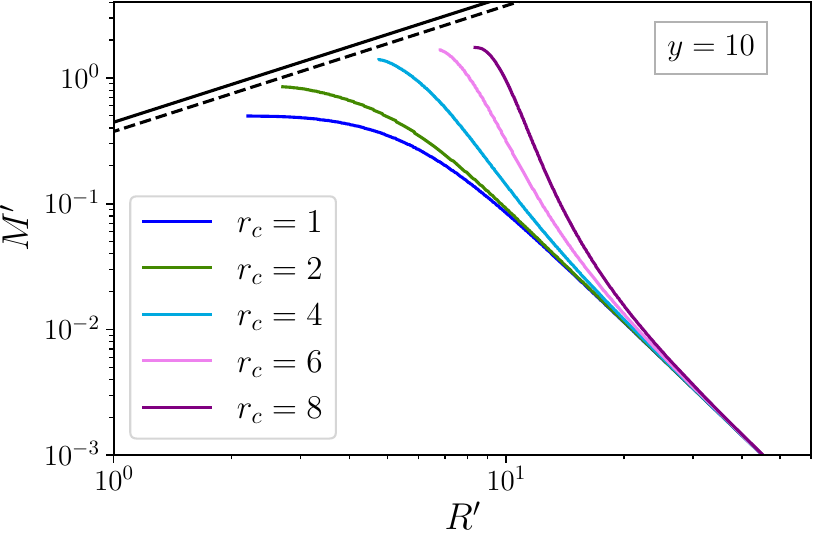}
		\includegraphics[width=0.45\textwidth]{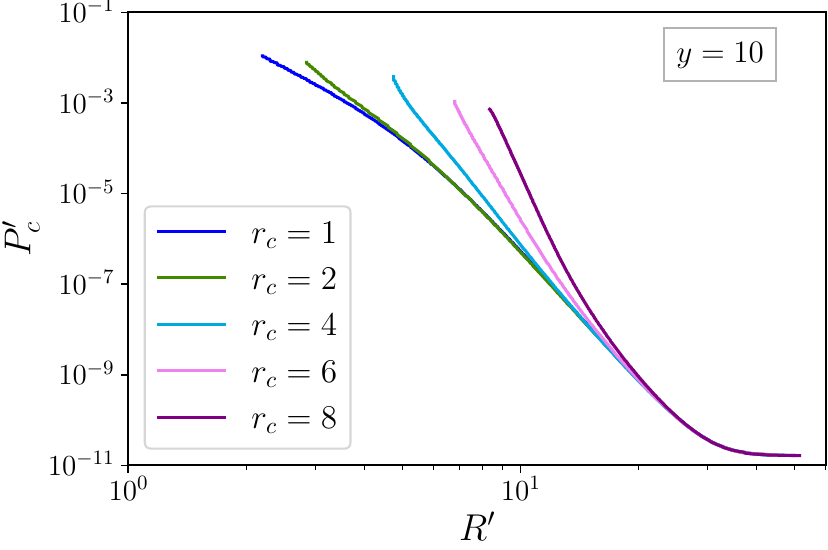}
	\caption{The dimensionless maximal mass $M'$ as a function of the radius $R'$ for different initial central pressure $P'_c$ (left) and the dimensionless central pressure $P'_c$ as a function of the radius of the star $R'$ (right) for a star combined out of a Fermi gas $m_1$ with interactions ($y=10$) as a core up to $r'_c$ followed by a free Fermi gas of dark matter particles $m_2=10^{-6}m_1$. The solid black line represents the Buchdahl limit and the dashed one the causality limit.}
	\label{fermi_plot}
\end{figure}

In this model we consider a star with an interacting Fermi gas core and a free Fermi gas shell. The core consists of fermionic matter with particle mass $m_f=m_1$ and the shell of particles with mass $m_f=m_2$.
For this combination we use an interaction strength of $y=10$ and a particle-mass-ratio of
$m_2/m_1=10^{-6}$ where the mass of the first matter type $m_1$ is our scaling unit, i.e. the
nucleon-mass, and the mass of the particles in the shell $m_2$ is very small and could be
considered as axinos or some kind of warm dark matter. In Fig. \ref{fermi_plot} we present the
dimensionless maximal masses $M'$ and  total radii $R'$ for different central pressures $P_c$.
The mass-radius-relation is only displayed for the stable solutions up to the maximal mass of the star. 

We notice that for the variation of the particle-mass-ratio $m_2/m_1$ at a fixed interaction-strength $y$ the mass-radius-relation or behaviour of the central pressure almost doesn't change. The interaction parameter $y$ dominates and determines the behaviour as can  be seen from the theoretical expression (\ref{fermi_int_press_eq}).  

From the mass-radius-relation we find that for small matching radii $r_c$ the star almost behaves  as a free Fermi gas and the slope of the mass radius relation roughly corresponds to the one for an EOS with a polytropic index of $\Gamma=5/3$, i.e. to the one of a fermion star without interactions.
When increasing the core-radius $r_c$ the slope of the mass changes and we are able to obtain more compact configurations. For large values of $r_c$ the interacting Fermi gas is the dominant component in the star and we find a change in the slope of the mass-radius curve. Then the slope comes close to that of a relativistic Fermi gas with $\Gamma=4/3$.

%##########################################################################
%+++++++++++++++++++++++++++++++ Quark- fermi++++++++++++++++++++++++++++++

\subsection{Self-bound matter and Fermi gas star } 
\label{sec:quark_fermi}

\begin{figure}
	\centering
		\includegraphics[width=0.45\textwidth]{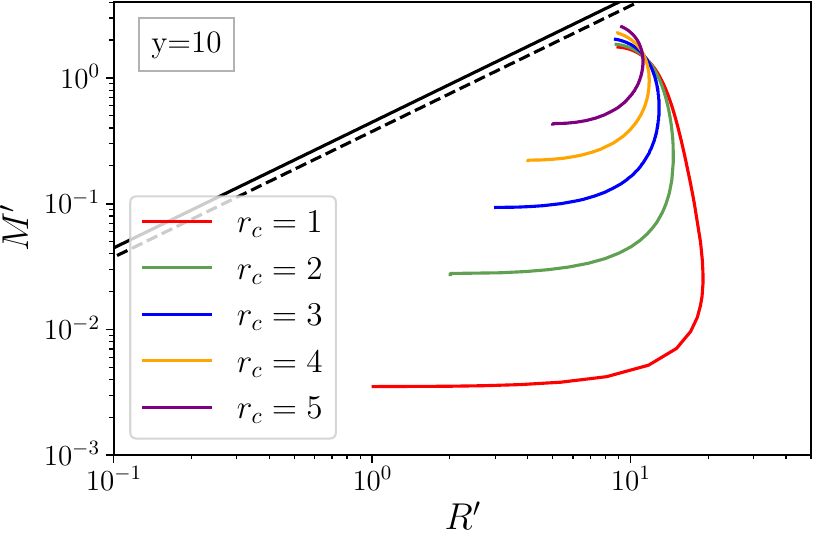}
     	\includegraphics[width=0.45\textwidth]{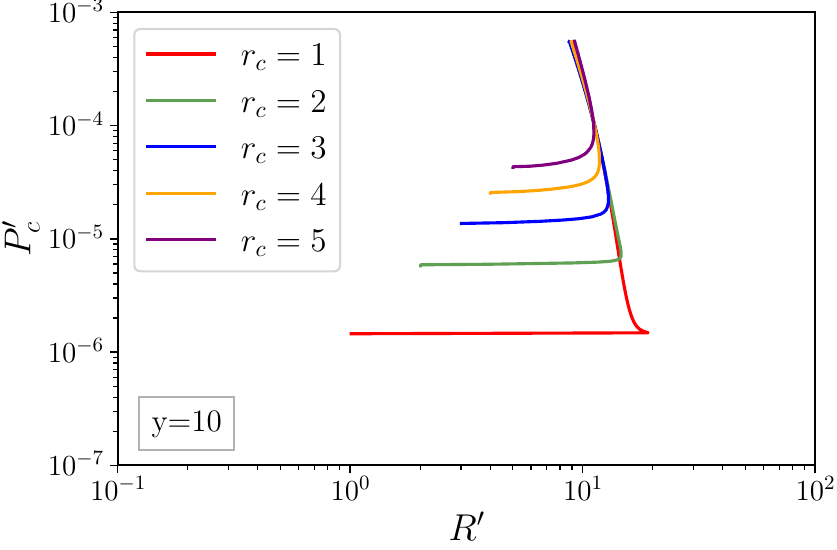}
	\caption{The dimensionless mass $M'$ (left) and pressure $P'$(right) as a function of the radius $R'$ for different initial values of the central pressure $P_c$ for a star combined out of a self-bound matter core and a Fermi gas with interactions ($y=10$) as a shell for different core-radii $r'_c$. Dashed and solid black lines are the causality and Buchdahl limit.  }
	\label{quark_plot}
\end{figure}

In this example we investigate a core described by the equation of state of self-bound matter, eq.~(\ref{eq:vacuumEOS}). The shell contains an interacting Fermi gas with particle mass $m_2=10^{-6}m_1$ and interaction strength $y=10$ which could represent some type of dark matter.  Again the first matter type $m_1$ is our scaling unit.
The core can be seen as dark matter coming from e.g.\ some dark QCD theory. Then some dark quarks are the components of the dark self-bound stars core.
In Fig. \ref{quark_plot} the dimensionless mass and pressure can be seen as a function of the total radius of the star for different core radii $r_c$. We present only the stable solutions of this configuration. The total mass $M'$ slowly increases for increasing central pressure $P'_c$ whereas the total radius $R'$ increases fast similar to the behaviour seen in model case two.
In this branch the star is dominated by the self-bound matter component then smoothly turning into the branch dominated by the interacting Fermi gas. There the total radius slowly decreases again for high pressures while the mass keeps increasing up to $M_{TOV}$. Afterwards the mass radius relation would produce a swirl due to the fermionic components of the shell resulting in an unstable branch which is not shown in Fig. \ref{quark_plot}. An unusual behaviour is found in the pressure for small core-radii as for the example of $r_c=1$. There the pressure increases very slow but the total radius of the star increases a lot and turns back in the direction of smaller radii for increasing central pressures. A similar behaviour has been seen for strange dwarfs, see
\citep{Glendenning:1994sp,Glendenning:1994zb} which consist of a core of absolutely stable strange quark matter and a halo of white dwarf matter. For a discussion of the stability analysis of strange dwarfs we refer to \citep{Alford:2017vca,DiClemente:2022ktz} noting that it does not directly apply to the case studied here with a constant core radius. 

%##########################################################################
%+++++++++++++++++++++++++++++ Mixed star++++++++++++++++++++++++++++++++++

\subsection{Mixed Fermi star} \label{sec:mixed}

\begin{figure}
\centering
\includegraphics[width=0.45\textwidth]{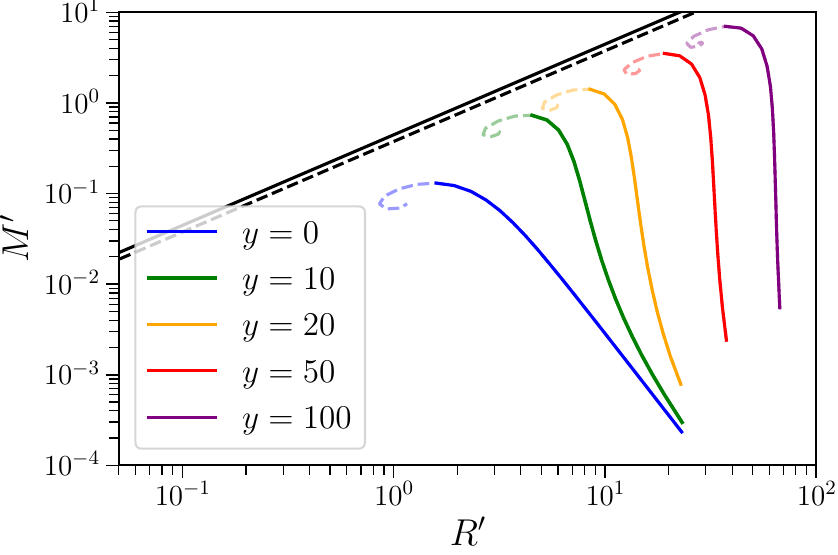}
\includegraphics[width=0.45\textwidth]{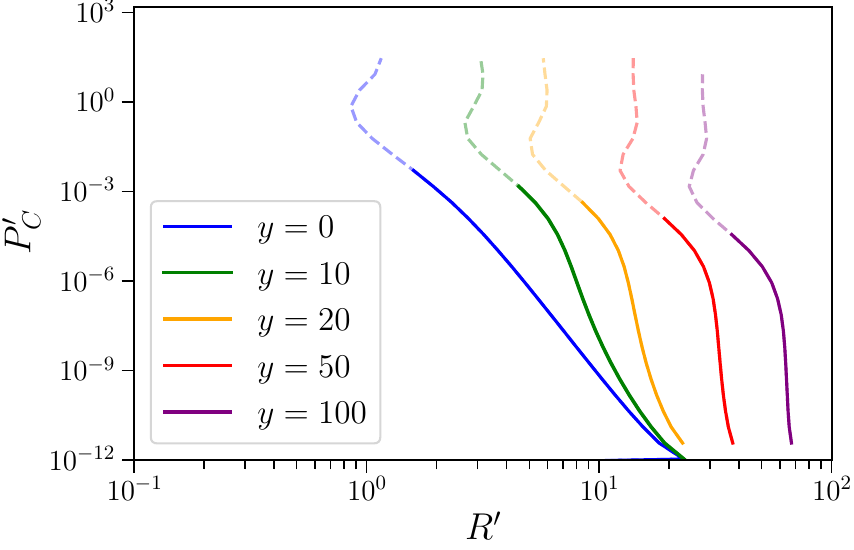}
\caption{The maximal mass (left) and central pressure (right) as a function of the total radius $R'$  for the mixed fluid containing an interacting Fermi gas with the particle-mass of $m_2 = 10^{-6}m_1$ and an interacting Fermi gas with particle mass $m_1$ and different interaction strengths $y =0, \, 10, \, 100, \, 1000 $ . The solid lines represent the stable branches while the dashed lines correspond to the unstable ones. The solid black line is the Buchdahl limit and the dashed black one is the causality limit.}
\label{mixeeed}
\end{figure}

Another case studied is a star with two homogeneously mixed components. 
The number densities of the two components are taken to be equal $n_1=n_2$. Such a case can be realized for the scenario of dark fermions with opposite dark charges, as studied for dark white dwarfs in \citep{Ryan:2022hku}.
In our case these two components are interacting Fermi gases, where one has a particle mass $m_1$ and a small particle mass $m_2=10^{-6}m_1$. 

In Fig. \ref{mixeeed} the mass-radius-relation and the pressure are presented in dimensionless units on a double-logarithmic scale. The mass radius relation reproduces the expected behaviour of a single interacting Fermi gas star. The stable branches with increasing mass up to $M_{TOV}$ can be well seen and the total mass of the star as well as $M_{TOV}$ increases for increasing interaction parameter $y$.
In this model, we kept the unstable solutions in the Fig.~\ref{mixeeed} (dashed lines) to show the typical behaviour of the Fermi gases and the similarities of the solutions for high central pressures. The mass radius relation - making a swirl at high central pressures -  corresponds to the wiggles seen in the pressure on the right.
Although the particle masses of the two components greatly differ, the behaviour is similar to that of a single interacting Fermi gas. Also, changing the ratio of the particle masses doesn't affect the behaviour since it is dominantly determined by the interaction strength.

%%%%%%%%%%%%%%%%%%%%%%%%%%%%%%%%%%%%%%%%%%%%%%%%%%%%%%%%%%%%%%%%%%%%%%%%%%%%%%

\subsection{Compactness}

\begin{figure}
	\centering
		\includegraphics[width=0.45\textwidth]{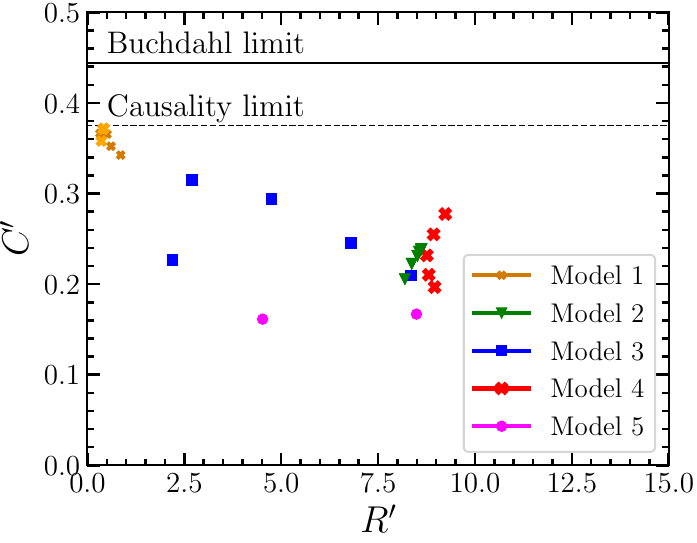}
	\caption{The compactness at $M_{TOV}$ the maximal mass for the five different models with different core radii $r_c$ and/or mass ratios.}
	\label{compactness}
\end{figure}

Fig. \ref{compactness} summarizes our results. There the dimensionless compactness $C$ is presented as a function of the total radius of the star $R'$ for all five model cases studied in this work. In several models we have investigated stars of two different types of matter combined at different values of the core radius $r_c$.
For each star with different core-radius we present here the compactness at the maximal total mass $M_{TOV}$ which corresponds to the most compact configuration for different $r_c$.

It can be seen that the stars combined of two incompressible fluids are the most compact ones.
As we have already observed before for very small core radii $r_c$ at a fixed density ratio  $\rho_1/\rho_0$, and large density ratios $\rho_1/\rho_0$ at fixed core radius the compactness almost reaches the causality limit of $C=3/8$.
It is a novel feature and finding that such a simple two-step-density profile configuration can reach such high values of the compactness. 

The second most compact combination is that of model case 3 where the star has an interacting Fermi gas core and a free Fermi gas shell consisting of very light particles.
This shows that even with a light component we can obtain compact configurations and we find that the behaviour is dominated more by the interaction strength than by the particle mass.

In the list of most compact configurations follows model case 4 which corresponds to a star containing a self-bound matter core and an interacting Fermi gas shell. The configuration with the largest amount of self-bound matter has the highest compactness. This model is followed by model 2 which corresponds to a a star with an incompressible fluid core and an interacting Fermi gas shell. The configuration with the smallest core radius is the most compact one.

The least compact configurations are obtained in model 5 in which the two interacting Fermi gases with a large particle mass ratio are mixed together. The mixed fluid is very insensitive to the particle mass but changes with the interaction strength of the Fermi gas as can be seen from the results in Figure \ref{mixeeed}.
Not all explored cases of the mixed star are shown in Fig. \ref{mixeeed}. The missing cases with a very high interaction strength of $y=50,\, 100$ have very large total radii of the star. We observe that the radius increases greatly with interaction strength, but the compactness stays almost the same and increases only slightly.

%%%%%%%%%%%%%%%%%%%%%%%%%%%%%%%%%%%%%%%%%%%%%%%%%%%%%%%%%%%%%%%%%%%%%%+++++++

\section{Summary} \label{sec:conclusions}
 
We have investigated 5 different model cases for the composition of exotic stars and obtained novel results for the mass-radius relation of compact stars made of fluids of dark matter. In the model cases we have combined two different types of matter with different particle mass ratios at different core-radii. We focused on scenarios with large particle mass ratios where the lighter component of the star can be thought as some sort of another dark matter particle. Such a light component, for example axinos, and the core-shell-structure of the star have an impact on the total mass and the compactness of the star. 

We find that the mass-radius relation for exotic compact objects with two different fluids of dark matter can exhibit
unusual and sophisticated patterns.
For comparison with our results specific for each model studied in this work listed below
we note that the characteristic mass-radius relation for white dwarfs and neutron stars follows 
usually simple typical relations.
For white dwarfs and low mass neutron stars the mass simply decreases with the radius cubed as dictated from the properties of a free Fermi gas equation of state. 
Single fluid selfbound stars, for an equation of state with a vacuum term, show as a simple
characteristic an increasing mass with radius cubed. 
 
First we have considered a model with two incompressible fluids with different density-ratios and different core radii which correspond to a two-step density profile. There we have obtained exotic mass-radius relations and 
find extremely compact configurations almost reaching the causality limit.
We have investigated the case in which we add an interacting Fermi gas shell on the core of an  incompressible fluid. The Fermi gas has particles with mass $m_f$ and is strongly interacting. We have found interesting shapes of the mass-radius-relation with very small total radii and masses for small central pressures. After the maximal total radius, the branch dominated by the incompressible fluid then turns to a branch dominated by the interacting Fermi gas which shows the typical behaviour of a Fermi gas.
We have found that increasing the core radius, i.e. the amount of incompressible fluid, the maximal mass of the star $M_{TOV}$ decreases. 
This happens due to the larger amount of the non-interacting incompressible fluid leading to less compact configurations compared to a simple interacting Fermi gas star.

The third model case we have investigated contains two Fermi gases. The core has components with a particle mass of $m_1$ and strong interactions. The shell is constructed of non-interacting particles with a small mass $m_2=10^{-6}m_1$. Such a combination of the different components leads to typical behaviours of the compact star configuration related to the Fermi gas in the relativistic and non-relativistic limit. It is new that we obtain configurations with a high compactness for some values of the core radius $r_c$. The total radius of the star is decreasing for an increasing value of the central pressure. 

In model case four we investigated a star with a self-bound matter core, described by a nonvanishing vacuum energy density and an interacting Fermi gas shell with strength $y=10$.
For small central pressures the star is dominated by the self-bound component and the total radius increases very fast while the central pressure and total mass increase slowly. This behaviour is similar to model case two in which for small central pressures the star is dominated by the component of the core and the central pressure looks almost horizontally for small values. A similar behaviour has already been seen in strange dwarfs. At some larger total radius of the star the total mass and pressure turn to smaller radii and increase in value. There the star is dominated by the Fermi gas. 

The last model case explored consists of two interacting Fermi gases with different interaction strength mixed together. One component has a particle mass of $m_1$ and the second component of the star in contrast has a small particle mass $m_2=10^{-6}m_1$. We have found that adding a component with such a small particle mass still reproduces the typical known behaviour of a Fermi star with an interacting Fermi gas. The total mass of the star and $M_{TOV}$ increases with increasing interaction strength. 

All the cases considered here lead to quite compact configurations. The cases containing two incompressible fluids give the most compact configurations, even getting very close to the causality limit. 

In summary, we can say that combining two different types of matter where either one or both of them could be dark matter gives interesting configurations in mass and radius when compared to known single fluid dark matter models. We even find stable solutions for such configurations. The additional dark matter component has an impact on the total mass and the total radius of the star, therefore making more compact configurations possible. The mass-radius relation can be substantially changed in some cases studied leading to unique features distinctly different to the mass-radius relation for ordinary matter, such as neutron stars and white dwarfs. 
We point out that this is an exploratory work for the study of two-component dark matter stars
which are encouraging. Certainly, the whole plethora of combinations of two dark matter fluids in exotic compact objects 
has just been touched upon.
Further investigations of exotic compact objects with different types of dark matter will be considered in future work.

%%%%%%%%%%%%%%%%%%%%%%%%%%%%%%%%%%%%%%%%%%%%%%%%%%%%%%%%%%%%%%%

\section{Conclusions}
From this study we conclude that it is possible to obtain different configurations of exotic stars containing dark matter and even two types of dark matter. This includes incompressible fluids, selfbound matter and fermionic dark matter.
The additional dark matter component leads to very compact configurations and also to new mass-radius curves which differ from those of stars without dark matter.
In models two (incompressible fluid core and interacting fermi gas shell) and four (selfbound matter core and interacting fermi gas shell) we obtain a large range of radii with almost the same total mass at small central pressures. 
Due to this property in the mass-radius-relation it would be possible to study mergers of dark matter stars with almost the same mass (mass ratio $q=1$) and therefore almost the same chirp mass, but with different radii. It is important to point out that the relic of the merger will not be a black hole but a stable compact star which will puff up to a considerably increased radius.
This would lead to an asymmetric gravitational wave signal which would differ from a signal of stars consisting of only one type of dark matter or of baryonic matter. 
This signal could be observed with the gravitational wave interferometers adLIGO, adVIRGO, LISA, the Einstein Telescope (ET) or the Big Bang Observer (BBO) and would be a probe of stars with dark matter, thus giving a constraint on the dark matter properties.
A sizeable difference should also be seen in the tidal deformability which scales with $R^5$ since two stars with almost the same mass, but different radii would give a small tidal deformability $\Lambda_{1}$ for the smaller star and a large $\Lambda_2 $ for the larger star. 
It is a novel result that we even obtain stable solutions for such exotic configurations and that these configurations may have a high compactness. 
Boson stars with the compactness of around
$ C  = 0.22$ are less compact than configurations with two types of dark matter from our study as in the models 1 to 4. The compactness is so high that it is more like that of black hole mimickers.

%%%%%%%%%%%%%%%%%%%%%%%%%%%%%%%%%%%%%%%%%%%%%%%%%%%%%%%%%%%%%%%%%%%%%%%%%%%%%%

\begin{acknowledgments}

This work emerged as a follow-up from the German-Canadian student research collaboration in theoretical astroparticle physics 'EXPeriential Learning Opportunity through Research and Exchange' EXPLORE of the York University, the University of Toronto, and the Goethe University of Frankfurt in the summer term of 2021. We gratefully thank 
Nassim Borzognia, Saeed Rastgoo, Laura Sagunski, Sean Tulin for initiating and setting up this research collaboration 
and all participants of EXPLORE for discussions and a fruitful collaboration. We thank Sean Tulin for a careful reading of the manuscript. We acknowledge support
by the Academic Innovation Fund of the York University, Toronto, Canada,
by the QSL (Quality Assurance in Teaching) of Goethe University Frankfurt,
by the Research Cluster 'ELEMENTS: Exploring the Universe from Microscopic to Macroscopic Scales' of the State of Hesse, 
and by the Deutsche Forschungsgemeinschaft (DFG, German Research Foundation) through the CRC-TR 211 ’Strong-interaction matter under extreme conditions’ -- project number 315477589-TRR-211.

\end{acknowledgments}

\bibliography{references.bib}
\bibliographystyle{aasjournal}

\end{document}